\documentclass[aps,showpacs,a4paper,superscriptaddress,prb]{revtex4-1}
\usepackage[latin9]{inputenc}
\usepackage[a4paper]{geometry}
\usepackage{amsmath}
\usepackage{amssymb}
\usepackage{graphicx}
\usepackage{esint}
\usepackage{color}

\usepackage{morefloats}

\DeclareMathOperator{\sign}{sign}
\renewcommand{\vec}[1]{\boldsymbol{#1}}

\begin{document}

\title{Interaction between point charges, dipoles and graphene layers}

\author{Francisco Guinea }

\email{Francisco.Guinea@Manchester.ac.uk}

\affiliation{Theoretical Physics Division, School of Physics and Astronomy, University
of Manchester, Manchester M13 9PL, UK}

\affiliation{IMDEA Nanoscience, C/Faraday, 9 Ciudad Universitaria de Cantoblanco
28049, Madrid, Spain}

\author{Niels R. Walet}

\email{Niels.Walet@Manchester.ac.uk}

\affiliation{Theoretical Physics Division, School of Physics and Astronomy, University
of Manchester, Manchester M13 9PL, UK}
\begin{abstract}
We analyse the interaction between charges and graphene layers. The electric polarisability of graphene
induces a force, that can be described by an image charge.  The analysis shows that graphene can be described as
 an imperfect conductor with a finite dielectric constant, $\epsilon_r$, for weak coupling, and it behaves as a metal for strong fields. 
As a consequence, the interaction between polar molecules and  graphene layer(s)
tends to align the molecular dipole along the
direction normal to the graphene, and quantitative estimates of the energy gain are given
The strength of this interaction can be
sufficient to overcome thermal effects when the molecule is close to
the layer, even at room temperature. Hence, boundary effects play a
significant role in determining the structure of systems such as water
confined in atomically narrow van der Waals heterostructures.
\end{abstract}


\maketitle

\global\long\def\ar{x_{\text{RPA}}}
\global\long\def\vF{v_{\text{F}}}
\global\long\def\aF{\alpha_{\text{G}}}
\global\long\def\atf{y_{\text{TF}}}

\section{Introduction}
Advances in the design of van der Waals heterostructures \cite{GG13}
have lead to the confinement of a number of substances between
graphene layers
\cite{Nair2012,algara-siller_square_2015,NPGG16,KGFGG16}. The materials
between the layers can experience large pressures, which can lead to
changes in their structure. The properties of matter under such extreme
pressure has been extensively studied using techniques such as the
Density Functional Theory \cite{MRMN03,SSOG04}, DFT, which can give a
good approximation to the ground state energy, and makes it possible to
compare different structures.  As a recent example,
consider the description of superconducting hydrogen sulfide under pressure,
e.g., Ref.\cite{duan_pressure-induced_2014,bernstein_what_2015,errea_high-pressure_2015}, where density functional theory is used extensively to describe 
the behaviour of bulk systems.

Most theoretical work on matter under high pressure has focused on
such bulk properties, without considering the role played by the boundary
conditions at the surface of the system.  Matter confined in van der
Waals heterostructures is in close contact with the layers in which it
is embedded. The influence of the boundary conditions on the total
energy is, most likely, comparable to or even exceeds the bulk
contribution. Moreover, graphene is a semimetal, which can interact
with electric charges and dipoles in the confined material. This gives
rise to image charge forces that are difficult to implement in
standard DFT calculations, although simple electrostatic arguments can
give a reasonable estimate. For example, most recent computational work on water
confined between graphene uses a simple surface integrated van der
Waals (3-9) force to describe the water-graphene interaction \cite{chen_two_2016,corsetti_structural_2016}, which is, at best, a crude approximation, since
it ignores the image charges caused by the semi-metal.
There
is substantial work where the water-graphene interaction is described
fully microscopically
\cite{freitas_dft_2011,ma_adsorption_2011,voloshina_physisorption_2011,li_influence_2012,wu_graphitic_2013,mckenzie_squeezing_2014},
but such approaches, although very insightful, seem of limited
practical use in complex calculations. It would thus be quite interesting
to see whether we can find a better effective model for the long-range
behaviour of the interaction between charges and graphene.

In the following, we focus on the dipolar interactions between water
molecules and graphene layers.  There are approaches where the
graphene and water are both described by density-functional theory
with van der Waals dispersive forces, e.g.,
Ref.~\cite{partovi-azar_many-body_2016}. As such calculations are very
complex, and thus have only limited reach, we would like to generate
models that include such interactions in a more approximate way, so
that we can concentrate on the behaviour of the confined molecules
without the additional complication of having to describe the detail
of the graphene.  To that end we study the effect of charges on
graphene layers and bilayers using two approximations, linear response and the Thomas-Fermi method.  
These methods are quite general, and can be applied to other dipolar molecules
and confining layers.  We first study the screening of a point charge
and a dipole outside a single graphene layer. The results are
generalized to a dipole between two layers in section
\ref{sec:bilayer}. In section \ref{sec:results} we present
quantitative estimates for the interaction of a water molecule within
a graphene bilayer, as function of the orientation of the
molecule. Some details of the calculations are discussed in the
Appendix.

\section{Single (flat) layer}

We first consider the interaction of a point charge with a single
graphene layer. We assume that the charge is at a distance $z$ from
the layer. The potential induced by the charge on the graphene layer,
$V ( \vec{r}_\parallel , z ) = Z e^2 / \sqrt{ | \vec{r}_\parallel^2 +
  z^2}$ reduces to the Coulomb potential at distances $|
\vec{r}_\parallel | \gg z$. The screening of a point charge in
graphene was treated in detail in\cite{DM84}. The effective fine
structure constant of graphene, $\aF = e^2 / ( 4 \pi \epsilon_0 \hbar
v_F )=Z\alpha c/v_F$, allows us to define two regimes, depending on
whether $Z \aF \lesssim 1$ or $Z \aF \gtrsim 1$\footnote{As usual $\alpha$ is
  the dimensionless coupling constant of QED,
  $\alpha=e^{2}/(4\pi\epsilon_{0}\hbar c)\approx1/137$.}
\cite{FNS07,SKL07,PNN07,SKL07b,KUPGN12}. We treat separately the cases
$Z \aF \ll 1$ (perturbative regime) and $Z \aF \gg 1$ (strong coupling).

\subsection{RPA screening: Subcritical regime, $Z\aF\ll1$}
\subsubsection{Point charge}
In the subcritical regime, where the effective coupling parameter
$Z\aF\ll1$, we
can use linear response to find the additional potential induced by
a charge placed near a graphene layer. Specifically, consider a point
charge $Ze$ displaced with respect to the origin a distance $\vec{r}_{\parallel0}$ parallel to the layer, and a distance
$z_{0}$ form the layer in the perpendicular direction. We assume that the layer
is infinitesimally thin, so that its polarisability can be described
by the equation
\begin{equation}
\chi(\vec{r}_{\parallel},z)=\chi(\vec{r}_{\parallel})\delta(z),
\end{equation}
where we assign the graphene layer the coordinate $z=0$. The potential
inside the graphene layer by the charge is $V_{0}(\vec{r}_{\parallel},0)=V_{\text{Coulomb }}(\vec{r}_{\parallel}-\vec{r}_{\parallel0},z_{0})$,
but this is screened due to the induced charge density inside the
layer,
\begin{equation}
V_{\text{scr}}(\vec{r}_{\parallel},z)=\frac{e^{2}}{4\pi\epsilon_{0}}\int d^{2}\vec{s}\frac{1}{\sqrt{|\vec{r}_{\parallel}-\vec{s}|^{2}+z^{2}}}\chi(\vec{r}_{\parallel}-\vec{s})V_{\text{tot}}(\vec{s},0),
\end{equation}
where $V_{\text{tot}}=V_{0}+V_{\text{scr}}$. This can easily be evaluated
in the von Laue coordinates $(\vec{q}_{\parallel},z)$
\begin{equation}
V_{\text{scr}}(\vec{q}_{\parallel},z)=\frac{1}{2\epsilon_{0}q_{\parallel}}e^{-q_{\parallel}z}\chi(\vec{q}_{\parallel})V_{\text{tot}}(\vec{q}_{\parallel},0).
\end{equation}
The polarisability of graphene in the RPA approximation is \cite{GGV99}
\begin{equation}
\chi(\vec{q}_{\parallel})=-\frac{n_{f}}{32}\frac{q_{\parallel}}{\hbar v_{F}},
\end{equation}
where $v_{F}$ is the Fermi velocity in the graphene layer, and $n_{f}=4$
is the number of fermion flavours in graphene. Making a 2D Fourier
transform of the Coulomb potential, we find
\begin{align}
V_{0}(\vec{q}_{\parallel},0) & =V_{\text{Coulomb}}(\vec{q}_{\parallel},z_{0})\nonumber \\
 & =Ze^{2}\frac{1}{2\epsilon_{0}q_{\parallel}}e^{i\vec{q}_{\parallel}\cdot\vec{r}_{\parallel0}}e^{-q_{\parallel}|z_{0}|}.
\end{align}
Assuming $z_{0}>0$, we now solve for the in-layer screening potential
$V_{\text{scr}}(\vec{q}_{\parallel},0),$ using the short-hand $\ar=e^{2}/16\epsilon_{0}\hbar v_{F}=\frac{\pi}{4}\aF$,
\begin{align}
V_{\text{scr}}(\vec{q}_{\parallel},0)\left(1-\frac{1}{2\epsilon_{0}q_{\parallel}}\chi(\vec{q}_{\parallel})\right) & =\frac{1}{2\epsilon_{0}q_{\parallel}}\chi(\vec{q}_{\parallel})V_{\text{0}}(\vec{q}_{\parallel},0),\nonumber \\
V_{\text{scr}}(\vec{q}_{\parallel},0) & =-\frac{\ar}{1+\ar}Z\frac{e^{2}}{2\epsilon_{0}q_{\parallel}}e^{i\vec{q}_{\parallel}\cdot\vec{r}_{\parallel0}}e^{-q_{\parallel}z_{0}}.
\end{align}
We can use this in turn to work out the general potential,
\begin{align}
V_{\text{scr}}(\vec{q}_{\parallel},z) & =\frac{1}{2\epsilon_{0}q_{\parallel}}e^{-q_{\parallel}z}\chi(\vec{q}_{\parallel})V_{\text{tot}}(\vec{q}_{\parallel},0)\nonumber \\
 & =e^{-q_{\parallel}z}V_{\text{scr}}(\vec{q}_{\parallel},0).
\end{align}
Thus, finally, transforming to coordinate space,
\begin{align}
V_{\text{scr}}(\vec{r}_{\parallel},z) & =\frac{\ar}{1+\ar}\frac{Ze^{2}}{4\pi\epsilon_{0}}\int dq_{\parallel}J_{0}(q|\vec{r}_{\parallel}-\vec{r}_{\parallel0}|)e^{-q_{\parallel}(z_{0}+z)}\nonumber \\
 & =\frac{\ar}{1+\ar}\frac{Ze^{2}}{4\pi\epsilon_{0}}Ze^{2}\frac{1}{\left(|\vec{r}_{\parallel}-\vec{r}_{\parallel0}|^{2}+(z+z_{0})^{2}\right)^{1/2}}.
\end{align}
This is just the effect of an imperfect image charge: in the metallic
limit ($\ar\rightarrow\infty$) we find that the screening potential
is equal but opposite to the applied potential, and in the vacuum
limit ($\ar=0$) we have no screening at all. Thus, using the results
in, e.g., Ref.~\cite{jackson1998classical}, we see that we can identify
\begin{equation}
\frac{\ar}{1+\ar}=\frac{\epsilon_{r}-1}{\epsilon_{r}+1},
\label{eq:areps1}
\end{equation}
or more simply
\begin{equation}
\epsilon_{r}=1+2\ar.
\label{eq:areps2}
\end{equation}

\subsubsection{Point dipole}

Since the dominant interaction of a polar but electrically neutral
molecule placed in front of a graphene will be caused by the dipole
force it is interesting to look at the simplified case of a point dipole.

We consider such a point-dipole  with  dipole moment $D$ at the point
$r_{\parallel}=0$, $z=z_{0}$, and the dipole moment making an angle $\theta$
with respect to the normal to the graphene layers.
The effect of the screening interaction
leads to an effective potential energy for the dipole, which orients
the dipole perpendicular to the surface,
\begin{align}
V_{\text{Dipole}} & =-\frac{\epsilon_{r}-1}{\epsilon_{r}+1}\frac{D^{2}(1+\cos^{2}\theta)}{32\pi\epsilon_{0}z_{0}^{3}}.
\end{align}
This agrees with the qualitative statement (see, e.g.,
Ref.\cite{partovi-azar2015vander}) that water molecules orient
preferentially perpendicular to the graphene layers.

\subsection{Strong coupling}

The RPA analysis, based on linear response theory, is valid for $Z\aF\ll1$.
The effect of a charged impurity, $Ze$, on a graphene layer shows
a transition to a strong coupling regime, the supercritical regime,
for $Z\aF\sim1$\cite{FNS07,PNN07,SKL07,SKL07b,KUPGN12}. In this
regime, the induced potential creates resonances within the graphene
band, and linear response theory is no longer applicable.

Let us first consider the extreme case $Z\aF\gg1$. The potential near
the external charge changes slowly with distance over distances
$r\sim d$. The induced charge density in the graphene layer, $\rho\sim
Z\aF/d^{2}$, also changes slowly with $r$, and leads to a screening
length\cite{FNS07} $k_\text{TF}^{-1}\sim d/(Z\aF)\ll d$.  This allows us to use
the Thomas-Fermi approximation\cite{DM84,BF09}.  In this
approximation the total energy of
the graphene layer can be expressed in terms of the induced electron
number density $\rho(\vec{r})$. This is defined relative to the background
density, so this can be positive or negative, but we find
that $\sign(Z)\rho$ is positive. 
Thus
\begin{align}
E_{\text{TF}} & =E_{\text{kin}}+E_{\text{pot}}+E_{\text{int}},\nonumber \\
E_{\text{kin}} & =\frac{2\hbar v_{F}}{3\pi}\int_{0}^{\infty}2\pi rdr\left[-\sign(Z)\pi\rho(r)\right]^{3/2},\nonumber \\
E_{\text{pot}} & =\frac{e^{2}}{4\pi\epsilon_{0}}\int_{0}^{\infty}2\pi rdr\frac{Z}{\sqrt{d^{2}+r^{2}}}\rho(r)\nonumber, \\
E_{\text{int}} & =\frac{e^{2}}{8\pi\epsilon_{0}}\int d^{2}\vec{r}\int d^{2}\vec{r}'\frac{\rho(\vec{r})\rho(\vec{r}')}{|\vec{r}-\vec{r}'|}.
\end{align}
It is convenient to express  the induced charge density as
\begin{align}
\rho(\vec{r}) & \equiv\frac{\tilde{\rho}\left(r/d\right)}{d^{2}},
\end{align}
where $\tilde{\rho}(x)$ is a dimensionless function. Then the Thomas-Fermi
approximation to the energy can be written as \cite{FNS07}
\begin{align}
E_{\text{TF}} =\frac{2\pi e^{2}Z}{4\pi\epsilon_{0}d}\biggl[& \frac{2}{3\pi Z\aF}\int_{0}^{\infty}xdx\left(-\sign(Z)\pi\tilde{\rho}(x)\right)^{3/2}\nonumber \\
 & +\int_{0}^{\infty}dx\frac{x\tilde{\rho}(x)}{\sqrt{1+x^{2}}}+\nonumber \\
 & +\frac{1}{2Z}\int_{0}^{\infty}xdx\int_{0}^{\infty}x'dx'\tilde{\rho}(x){\cal F}(x,x')\tilde{\rho}(x')\biggr].\label{eq:ETF}
\end{align}
Here
\begin{align}
{\cal F}(x,x') & =\frac{E_{K}\left[\frac{4xx'}{(x+x')^{2}}\right]}{x+x'},
\end{align}
and $E_{K}(x)$ is the complete elliptic integral of the first kind.
Ignoring the overall scale of the energy, we realise that the kinetic energy is a small perturbation to the energy. We thus write a perturbation expansion for
\begin{equation}
\tilde{\rho}(x)=\tilde{\rho}_{0}(x)+\frac{1}{\aF Z}\tilde{\rho}_{1}(x)+\mathcal{O}\left(\frac{1}{(\aF Z)^{2}}\right).
\end{equation}
We substitute this in the Euler-Lagrange equation, and
we find an exact solution for $\rho_0$,
\begin{align}
\tilde{\rho}_{0}(x) & =\frac{-Z}{2\pi(1+x^{2})^{3/2}},
\end{align}
which corresponds to a perfect (metallic) image charge. This is due to the fact
that have used Coulomb's
law with $\epsilon=\epsilon_{0}$ in the final two terms of Eq.~(\ref{eq:ETF}).

The next correction is given by the solution to the Fredholm equation
\begin{equation}
|Z|\left(\frac{\pi|Z|}{2\pi(1+x^{2})^{3/2}}\right)^{1/2}=-\frac{1}{Z}\int_{0}^{\infty}x'dx'\mathcal{F}(x,x')\tilde{\rho}_{1}(x'),
\end{equation}
which can probably no longer be solved analytically.
Since  the Thomas-Fermi approximation is only exact to the lowest order, we need to be slightly suspicious about solutions beyond this regime.
Nevertheless,  one way to make progress is to deal with $\rho_1$, etc.,
approximately by employing a variational Ansatz with an effective screening factor $Z_\text{TF}$,
\begin{align}
\rho(x) & =\frac{-Z_{\text{TF}}}{2\pi(1+x^{2})^{3/2}},
\end{align}
which gives
\begin{align}
E_{\text{TF}}(Z_{\text{TF}}) & =\frac{\pi e^{2}}{4\pi\epsilon_{0}d}\left[\frac{2\sqrt{2}}{15\pi\aF}\left|Z_{\text{TF}}\right|^{3/2}+ZZ_{\text{TF}}+\frac{Z_{\text{TF}}^{2}}{2}\right].
\end{align}
This energy is minimised for
\begin{align}
\sqrt{-Z_{TF}/Z}=-\atf+\sqrt{\atf^{2}+1}
\end{align}
with
\begin{align}
\atf & =\frac{\sqrt{2}}{10\pi\aF\sqrt{Z}}
\end{align}
Asymptotically, we find
\begin{align}
-\frac{Z_{TF}}{Z} & \approx1-2y_{\text{TF}}+\mathcal{O}\left(y_{\text{TF}}^{2}\right),\quad\atf\ll1.
\end{align}
Thus the Thomas-Fermi approximation gives results that are not that different from the
RPA approximation, but the effective behaviour has changed from semi-metallic (imperfect image charges) 
to metallic (perfect image charge).

This similarity between these two approach in 2D materials has
been observed before:  A related discussion regarding the similarity between
linear response and the Thomas-Fermi approximation can be found in the work by
Stott and collaborators \cite{koivisto2007kinetic,calderin2010performance}, who conclude that
two dimensions  the two are rather similar.

\section{Bilayer}
\label{sec:bilayer}

The effect of (a set of) charges between a graphene bilayer, similar
to the case of a parallel plate capacitor, is of particular interest
if we consider van der Waals sandwiches, where the two outer layers
are graphene. 
In this regime we should probably only consider the weak coupling, since we have no unscreened charges.
Of course, the Fermi velocity in these outer layers
could be rather different, e.g., if one is mounted on a substrate, and
the other not.

\subsection{Equal Fermi velocities}

Consider a point charge at position $\vec{r}_{\parallel0}$ parallel
and distance $z_{0}$ perpendicular in between two otherwise identical
graphene layers positioned at a distance $d$. We shall chose the
$z$ coordinate of the layers $\pm d/2$, respectively. The two graphene
layers experience an electrostatic potential $V_{0}(\vec{r}_{\parallel},0)=V_{\text{Coulomb }}(\vec{r}_{\parallel}-\vec{r}_{\parallel0},|z_{0}\pm d/2|)$
from this charge, which induces a charge density in the graphene layers
(again $V_{\text{tot}}=V_{0}+V_{\text{scr}}$). We assume that these
layers are thin, and use the linear response approach, thus
\begin{equation}
\chi(\vec{r}_{\parallel},z)=\chi(\vec{r}_{\parallel})\left(\delta(z-d/2)+\delta(z+d/2)\right).
\end{equation}
 We again work in von Laue coordinates and as above we evaluate the
screening potential in terms of the linear response to the total potential,
\begin{equation}
V_{\text{scr}}(\vec{q}_{\parallel},z)=\frac{e^{2}}{2\epsilon_0q_{\parallel}}\chi(\vec{q}_{\parallel})\left(e^{-q_{\parallel}|z-d/2|}V_{\text{tot}}(\vec{q}_{\parallel},d/2)+e^{-q_{\parallel}|z+d/2|}V_{\text{tot}}(\vec{q}_{\parallel},-d/2)\right).\label{eq:scr21}
\end{equation}

After some mathematical manipulations (see appendix) we find that, 
assuming $|z|,|z_{0}|\le d/2$,
\begin{align}
V_{\text{scr}}(\vec{q}_{\parallel},z) & =-\ar Z\frac{e^2}{2\epsilon_0q_{\parallel}}e^{i\vec{q}_{\parallel}\cdot\vec{r}_{\parallel0}}\biggl(e^{-q_{\parallel}(d/2-z)}\left[\delta c_{1}(q_{\parallel})e^{-q_{\parallel}(d/2-z_{0})}+c_{2}(q_{\parallel})e^{-q_{\parallel}(d/2+z_{0})}\right]\nonumber \\
 & \qquad+e^{-q_{\parallel}(d/2+z)}\left[\delta c_{1}(q_{\parallel})e^{-q_{\parallel}(d/2+z_{0})}+c_{2}(q_{\parallel})e^{-q_{\parallel}(d/2-z_{0})}\right]\biggr)\\
 & =Z\frac{e^2}{2\epsilon_0 q_{\parallel}}e^{i\vec{q}_{\parallel}\cdot\vec{r}_{\parallel0}}\frac{-\cosh(z+z_{0})+e^{-a-q_{\parallel}d}\cosh(z-z_{0})}{\sinh(a+q_{\parallel}d)}.
\end{align}
Comparing with equivalent expression for the Green's function of a
system with two layers with relativity permittivity $\epsilon_{r}$,
e.g. in Ref. \cite{Otani}, we see that this is once again correct if we
assume Eqs.\ (\ref{eq:areps1} and (\ref{eq:areps2}), and thus
\begin{equation}
 a=\ln\left(\frac{\epsilon_{r}+1}{\epsilon_{r}-1}\right).
\end{equation}

\section{Results}
\label{sec:results}

\begin{figure}
\begin{centering}
\includegraphics[width=10cm]{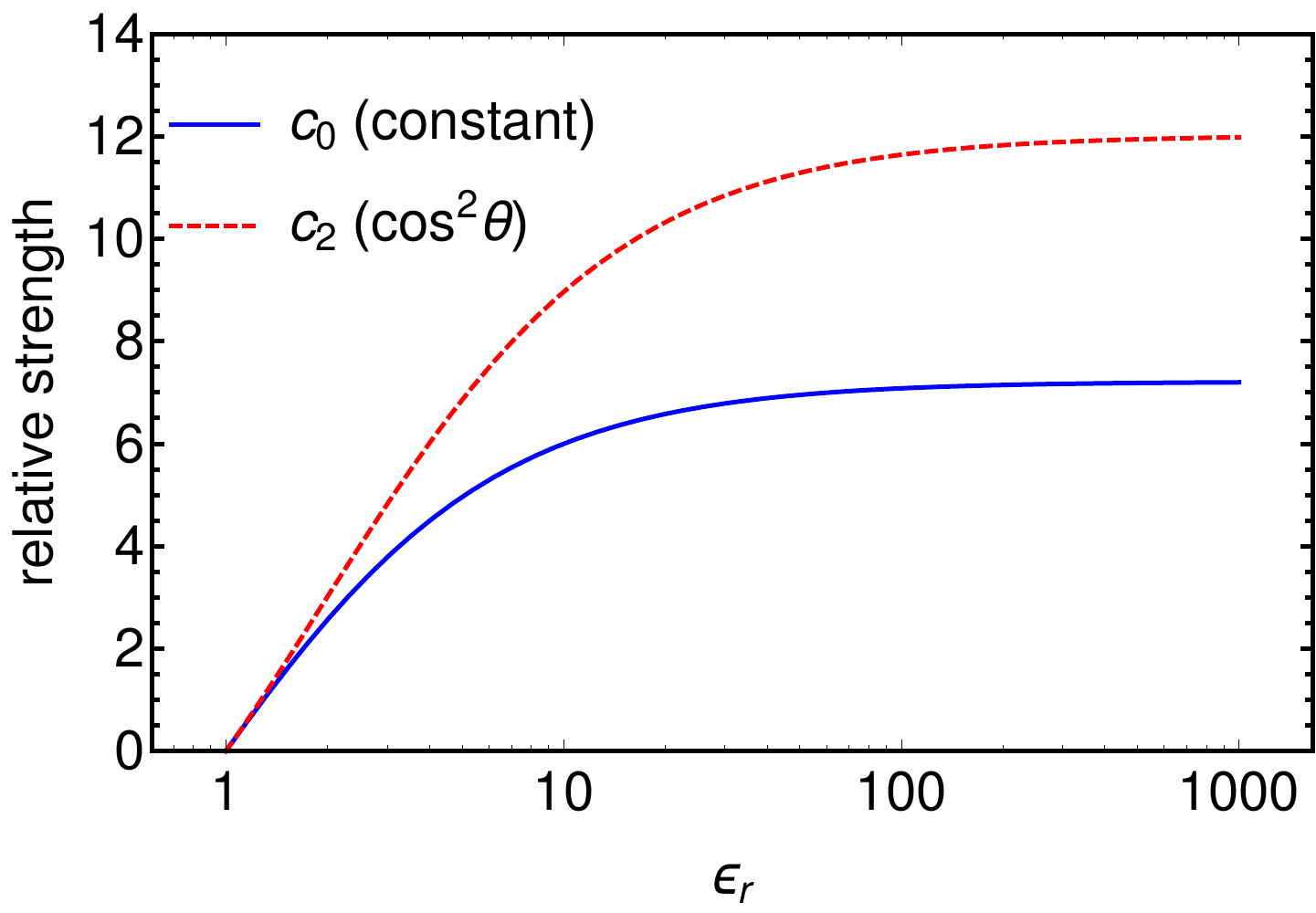}
\end{centering}
\caption{(Colour online) The behaviour of the coefficients of the constant and of the $\cos^2 (\theta)$ terms in Eq.~(\ref{eq:Ez0}) as a function of $\epsilon_r$.
\label{fig:The-two-coefficients}}
\end{figure}

We now once again look at the interaction of a point dipole placed
at vertical position $z$, with orientation $\theta$ relative to
the normal. Calculating the interaction energy as an integral over
$q_{\parallel}$, we find that
\begin{align*}
E(z,\theta) & =-\frac{1}{2}D^{2}\int dq_{\parallel}\,q_{\parallel}^{2}\text{csch}(a+dq_{\parallel})\left(\left(\cos^{2}\theta+1\right)\cosh(2zq_{\parallel})+e^{-a-dq_{\parallel}}\left(3\cos^{2}\theta-1\right)\right)\\
 & =-\frac{D^{2}}{4d^{3}}\biggl\{\frac{1}{2}\frac{\epsilon_r-1}{\epsilon_r+1}\left[\Phi\left(\left(\frac{\epsilon_r-1}{\epsilon_r+1}\right)^{2},3,\frac{1}{2}-\frac{z}{d}\right)+\Phi\left(\left(\frac{\epsilon_r-1}{\epsilon_r+1}\right)^{2},3,\frac{1}{2}+\frac{z}{d}\right)\right]\left(\cos^{2}\theta+1\right)\\
 & \qquad+\text{Li}_{3}\left(\left(\frac{\epsilon_r-1}{\epsilon_r+1}\right)^{2}\right)\left(3\cos^{2}\theta-1\right)\biggr\}.
\end{align*}
Here $\Phi ( z , s, a )$ is the Hurwitz-Lerch transcendental function, and $Li_3 ( z )$ is the polylogarithm function.  
This shows that it is energetically favourable for the dipole to be
perpendicular to the surface; also, the central position is an unstable
position, and the dipole would like to move close to the surface.
For a centrally placed dipole we find, by writing
\begin{equation}
E=-\frac{D^{2}}{16\pi\epsilon_0d^{3}}(c_0(\epsilon_r)+c_{2}(\epsilon_r)\cos^{2}\theta),\label{eq:Ez0}
\end{equation}
the results shown in Fig.~\ref{fig:The-two-coefficients}.

The prefactor $D^2/(16\pi\epsilon_0 d^3)$ takes the value $4.27\text{ meV}$ for water
(using $D=1.8546\pm0.0006\text{ Debye units (10$^{-18}$ esu cm}$) \cite{clough_dipole_1973}
and $d=5 \text{ \AA}$; The attraction associated with
perpendicular alignment is $51\text{ meV}$ in the metallic regime, and $29\text{ meV}$ for
$\epsilon_r=5$.

There are a couple of recent experimental results where the dipole
force will play an important role. The fist of these is the
confinement of water between to graphene layers
\cite{algara-siller_square_2015}, which will be discussed in more
detail in Ref.\ \cite{WaletWater}, since other effects are of real importance there. The second, which we will study here, is the flow of water
through graphene microchannels \cite{algara-siller_square_2015,NPGG16}. The analysis in that paper
suggests a flow that is very different for a very shallow channel; all
evidence suggests that in such channels we have an enhanced flow
through organised layers, rather than the more mixed Poisseuille flow
for deeper channels \cite{Jichenstudent}. That raises the question
whether the dipole force gives rise to more organisation in such
layers.

\begin{figure}
\begin{centering}
\includegraphics[width=10cm]{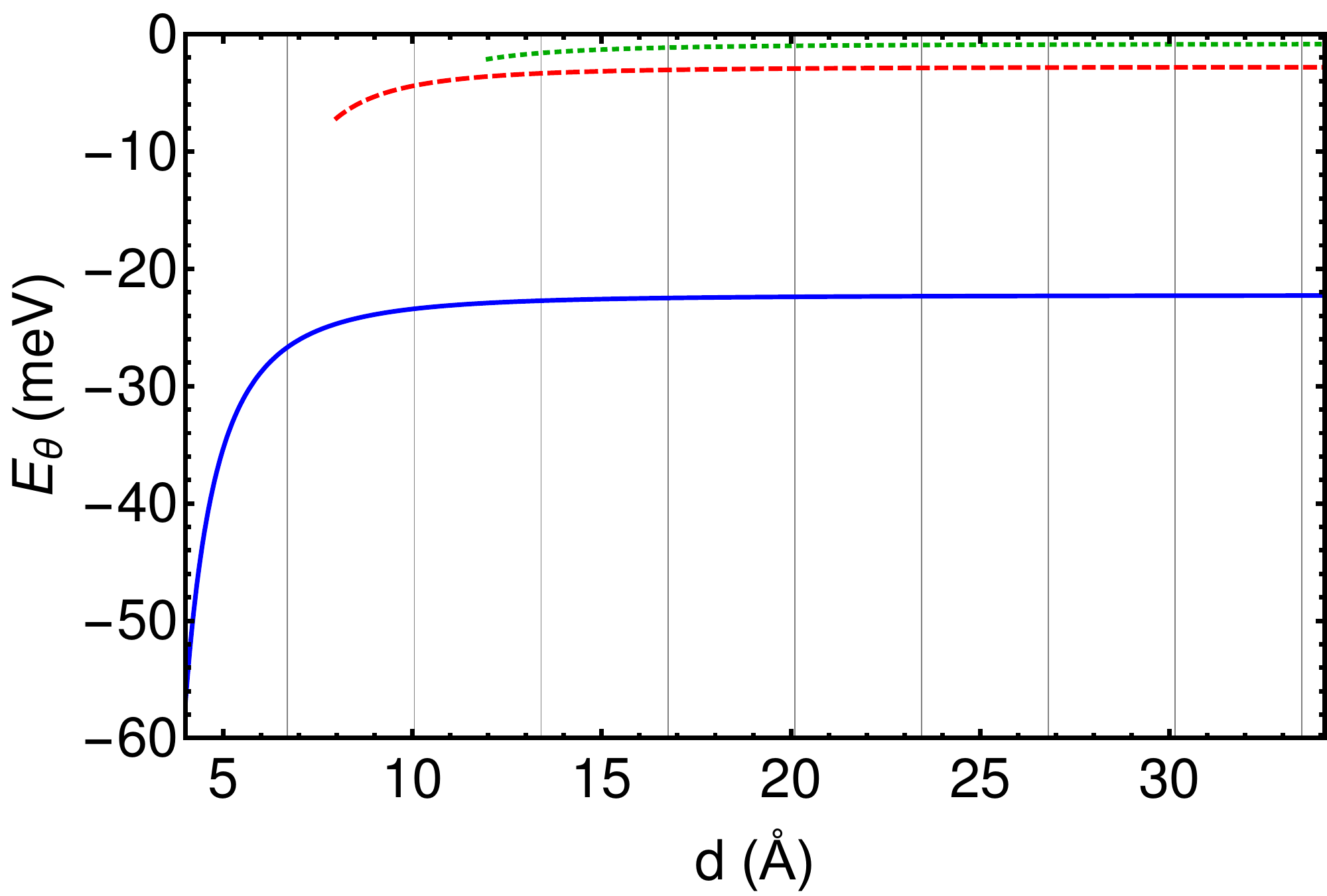}
\end{centering}
\caption{(Colour online) The energy gain  for   water 
approximated as a point dipole
placed in a perpendicular orientation in a channel between
two graphene layers, as a function of the channel height $d$. Results are shown
for three different distances between the center of the water molecule and one of the graphene layers (blue solid $z=2\text{ \AA}$, red dahsed $z=4\text{ \AA}$, green dotted $z=6\text{ \AA}$).
 The grid lines indicate probable values for the channel height, based on multiples the graphite layer spacing of $3.35\text{ \AA}$.}
\label{fig:orientation}
\end{figure}

Water will clearly orient if the interaction of a water molecule with the material is stronger
than the thermal energy. We see in
 Fig.\ \ref{fig:orientation}  that the outside layer of water
will most likely direct perpendicular to the graphene layers 
($kT\simeq 25\text{ meV}$); the other layers
will not orient due the interaction of individual water molecules with
the graphene, but may once we take into account the dipole
interactions between the water molecules in the layers as well. This suggest that it would be important to include the long-range electromagnetic interactions in the simulations. 

\section{Conclusions}
We have shown that boundary effects can be important in determining the properties of water and other polar molecules confined within graphene layers. We show that both in the weak-field linear response theory and in the strong-coupling regime we have to deal with image charges; the only difference is that
for strong fields graphene behaves as a metal. 

The interaction between the molecules and graphene can be approximated using an image charge model. We have estimated the strength of the image charges both in the weak and strong coupling limits. We have considered neutral graphene layers. The formalism can easily be extended to the case when graphene is charged. As the polarizability of the graphene layers will increase, our results can be considered a lower bound to the influence of boundary effects.

The graphene layer tends to orient the molecular dipoles in the direction normal to the layer. For molecules at a few angstroms of the layer, the energy associated to the oriented configuration is comparable, or higher, than room temperature. This energy, for the case of water, is also comparable to the molecule-molecule interaction, which, to a large extent, is also of electrostatic origin\cite{J81}. 

The combination of boundary effects and molecule-molecule interactions, for a single water layer embedded between graphene sheets, will most likely lead to an antiferroelectric arrangement, where the dipoles are oriented in the direction normal to the layers.

\section{Acknowledgements}
We acknowledge helpful conversations with M. A. Moore and A. K. Geim. F. G. acknowledges financial support from the European Research Council, grant 290846, the European Commission under the contract CNECTICT-
604391, the Graphene Flagship, and MINECO (Spain), grant FIS2014-57432.

\appendix
\section{Mathematical details}
\subsection{Derivation of potential for bilayers}
Following from Eq.\ (\ref{eq:scr21}), 
evaluating the screening potential inside the two layers gives rise
to two coupled equations
\begin{align}
V_{\text{scr}}(\vec{q}_{\parallel},\pm d/2) & =-\ar\left(V_{\text{scr}}(\vec{q}_{\parallel},\pm d/2)+V_{0}(\vec{q}_{\parallel},\pm d/2)+e^{-q_{\parallel}d}\left(V_{\text{scr}}(\vec{q}_{\parallel},\mp d/2)+V_{0}(\vec{q}_{\parallel},\mp d/2)\right)\right),
\end{align}
which we can solve easily:
\begin{align}
V_{\text{scr}}(\vec{q}_{\parallel},0\pm d/2) & =c_{1}(q_{\parallel})V_{0}(\vec{q}_{\parallel},\pm d/2)+c_{2}(q_{\parallel})V_{0}(\vec{q}_{\parallel},\mp d/2),\\
c_{1}(q_{\parallel}) & =-1+\delta c_{1}(q_{\parallel})\nonumber \\
 & =-1+\frac{e^{q_{\parallel}d}/\ar}{2\sinh(a+q_{\parallel}d)},\\
c_{2}(q_{\parallel}) & =-\frac{e^{-a}/\ar}{2\sinh(a+q_{\parallel}d)},
\end{align}
Here
\begin{equation}
a=\ln\left(\frac{1+\ar}{\ar}\right).\label{eq:a}
\end{equation}
Thus
\begin{align}
V_{\text{tot}}(\vec{q}_{\parallel},0\pm d/2) & =\delta c_{1}(q_{\parallel})V_{0}(\vec{q}_{\parallel},\pm d/2)+c_{2}(q_{\parallel})V_{0}(\vec{q}_{\parallel},\mp d/2)\,.
\end{align}
 From this we can reconstruct the full potential in momentum space
\begin{align}
V_{\text{scr}}(\vec{q}_{\parallel},z) & =-\ar\left(e^{-q_{\parallel}|z-d/2|}V_{\text{tot}}(\vec{q}_{\parallel},d/2)+e^{-q_{\parallel}|z+d/2|}V_{\text{tot}}(\vec{q}_{\parallel},-d/2)\right)\nonumber \\
 & =-\ar Z\frac{e^2}{2\epsilon_0q_{\parallel}}e^{i\vec{q}_{\parallel}\cdot\vec{r}_{\parallel0}}\biggl(e^{-q_{\parallel}|z-d/2|}\left[\delta c_{1}(q_{\parallel})e^{-q_{\parallel}|z_{0}-d/2|}+c_{2}(q_{\parallel})e^{-q_{\parallel}|z_{0}+d/2|}\right]\nonumber \\
 & \qquad+e^{-q_{\parallel}|z+d/2|}\left[\delta c_{1}(q_{\parallel})e^{-q_{\parallel}|z_{0}+d/2|}+c_{2}(q_{\parallel})e^{-q_{\parallel}|z_{0}-d/2|}\right]\biggr).
\end{align}

\subsection{Unequal Fermi velocities}
It is straightforward to generalise these results to two layers with
unequal Fermi velocities, e.g., one layer on a substrate and the other
free-standing.  We introduce the notation $x_\pm$ for the value of
$\ar$ for the upper ($+$) and lower ($-$) layer, and associate
parameters $a_\pm$ in a way similar to Eq.\ (\ref{eq:a}). We then
find
\begin{align}
V_{\text{scr}}(\vec{q}_{\parallel},0\pm d/2) & =c_{1\pm}(q_{\parallel})V_{0}(\vec{q}_{\parallel},\pm d/2)+c_{2\pm}(q_{\parallel})V_{0}(\vec{q}_{\parallel},\mp d/2),\\
c_{1\pm}(q_{\parallel}) & =-1+\frac{e^{x_\mp/2+q_{\parallel}d}\sinh(x_\pm/2)}
{\sinh((a_++a_-)/2+q_{\parallel}d)},\\
c_{2\pm}(q_{\parallel})&=-\frac{e^{x_\mp/2}\sinh(x_\pm/2)}{\sinh((a_++a_-)/2+q_{\parallel}d)}.
\end{align}

\bibliographystyle{aipnum4-1}
\bibliography{polarization_surface}

\end{document}